# Electronic Transport in Chemical Vapor Deposited Graphene Synthesized on Cu: Quantum Hall Effect and Weak Localization

*Helin Cao[1,2], Qingkai Yu[3,#], Luis A. Jauregui[2,5], Jifa Tian[1,2], Wei Wu[3], Zhihong Liu[3], Romaneh Jalilian[1,2], Daniel K. Benjamin[4], Zhigang Jiang[4], Jiming Bao[3], Steven S.S. Pei[3] and Yong P. Chen[1,2,5,\*]*

[1] Department of Physics, Purdue University, West Lafayette, IN 47907

[2] Birck Nanotechnology Center, Purdue University, West Lafayette, IN 47907

[3] Center for Advanced Materials, Department of Electrical and Computer Engineering, University of Houston, Houston, TX 77204

[4] School of Physics, Georgia Institute of Technology, Atlanta, GA 30332

[5] School of Electrical and Computer Engineering, Purdue University, West Lafayette, IN 47907

\# Email: qyu2@uh.edu,  \* Email: yongchen@purdue.edu

**Abstract**

**We report on electronic properties of graphene synthesized by chemical vapor deposition (CVD) on copper then transferred to $SiO_2$/Si. Wafer-scale (up to 4 inches) graphene films have been synthesized, consisting dominantly of monolayer graphene as indicated by spectroscopic Raman mapping. Low temperature transport measurements are performed on micro devices fabricated from such CVD graphene, displaying ambipolar field effect (with on/off ratio ~5 and carrier mobilities up to ~3000 $cm^2$/Vs) and "half-integer" quantum Hall effect, a hall-mark of intrinsic electronic properties of monolayer graphene. We also observe weak localization and extract information about phase coherence and scattering of carriers.**

Graphene, a single layer of graphite, has attracted tremendous interests as a novel electronic material with many potential applications[1,2]. The initial experiments revealing graphene's unique electronic transport properties (such as ambipolar field effect[3] and "half-integer" quantum Hall effect[4,5]) were performed with graphene mechanically exfoliated from graphite. While exfoliation typically gives only small (tens of μm) graphene flakes and is not a scalable method to produce graphene for practical applications, many other methods are developed to synthesize high quality graphene at large scale. One example is epitaxial growth on SiC[6,7]. Another example is chemical vapor deposition (CVD) on metals. Metal-based CVD is a decades-old method to grow graphene (see reviews in Refs. 8-9). Lately it has received revived interests as a potentially scalable way to produce graphene that can be readily transferred to other substrates for electronic applications[10-19]. In particular, copper (Cu) has been demonstrated as an exceptional metal substrate allowing CVD growth of large-size single layer graphene with excellent quality and uniformity[15-17] and promise for applications in transistors[15,18], transparent electrodes[17,19] and flexible electronics[17,19].

In this letter, we demonstrate that CVD graphene grown on Cu possesses intrinsic graphene electronic properties by observing the hall-mark half-integer quantum Hall effect (QHE) in low temperature magnetotransport measurements. We also study the weak localization to probe carrier scattering and phase coherence. Our results will be important for understanding the electronic properties of CVD-grown graphene on Cu and using such graphene in fundamental research or electronic applications.

The CVD-graphene samples used in this work are synthesized on Cu foils using procedures[20] analogous to those in published literature[15-19] and in our previous work[21]. The growth pressure is 1 atm (ambient pressure). Following the synthesis, the Cu is etched and graphene transferred (assisted by PMMA[11,15]) onto $SiO_2$/Si. Fig. 1a,b shows a large size (4 in × 4 in) graphene we synthesized. Fig. 1c shows representative Raman spectra measured (with a 532nm laser) at several different spots from the transferred film. The very low disorder-induced "D" band (~1350 cm$^{-1}$) indicates the high quality of the synthesized graphene[22]. The large intensity ratio ($I_{2D}/I_G$ > 2) of 2D band (~2680cm$^{-1}$) over G band (~1580 cm$^{-1}$) is associated with monolayer graphene[15,22]. One possible reason for the relatively large variation of $I_{2D}/I_G$ could be the spatially non-uniform adhesion between the transferred graphene and the substrate ($SiO_2$), as the substrate-interaction can strongly affect the Raman spectrum for monolayer graphene[23]. Fig. 1d shows a 200 μm ×200 μm Raman map of $I_{2D}/I_G$, with ~99% of the area having $I_{2D}/I_G$ >2 (and ~90% having $I_{2D}/I_G$ > 3), indicating the film is predominantly monolayer graphene.

The transferred graphene is fabricated into quasi-Hall-bar devices using e-beam lithography, $O_2$ plasma etching and metallization (evaporated Ti/Au contacts). The optical image of a representative device ("A") is shown in Fig. 2a inset. Electrical resistances are measured by low frequency, low current lock-in detection. The carrier density is tuned by a back gate voltage ($V_{gate}$) applied to the highly doped Si substrate, with the 300nm $SiO_2$ as the gate dielectric. Measurements on several such devices yield similar results. Data from two devices ("A" and "B") are presented below.

Fig. 2a shows 4-terminal resistance ($R_{xx}$) vs. $V_{gate}$ measured in device "A" at low temperature ($T$=0.6 K) and zero magnetic field. We observe the characteristic "ambipolar" field effect[3-5,11,12,14,15], with resistance modulation ratio more than 5. The charge neutral "Dirac point" ($V_{DP}$, position of resistance peak) is ~20V for this sample (the positive $V_{DP}$ indicates some extrinsic "residual" hole-doping, common in fabricated graphene devices[3]). The field effect mobility[3-5,11,12,14,15] is ~3000 $cm^2$/Vs for holes ($V_{gate}<V_{DP}$) and ~1000 $cm^2$/Vs for electrons ($V_{gate}>V_{DP}$). Similar field effect is also observed at room temperature, although we can access a larger range of $V_{gate}$ at lower $T$ without gate leakage.

Fig.2b shows $R_{xx}$ (4-terminal longitudinal resistance) and $R_{xy}$ (Hall resistance) of device "A" vs. $V_{gate}$ measured at a high magnetic field ($B$=18T, perpendicular to the sample) and low $T$ (0.7 K). The sign reversal of $R_{xy}$ from $V_{gate}>V_{DP}$ to $V_{gate}<V_{DP}$ is consistent with the ambipolar field effect (change of carrier types). Most remarkably, $R_{xy}$ is seen to exhibit clearly quantized plateaus at $h/(2e^2)$ for electrons, $-h/(2e^2)$, $-h/(6e^2)$ and $-h/(10e^2)$ for holes, all accompanied by vanishing $R_{xx}$, where $e$ is the elementary charge and $h$ is the Plank constant. Such a "half-integer" quantum Hall effect (QHE) is an electronic hall-mark of monolayer graphene[4,5]. The LL filling factor ($v$) for the observed QHE states in Fig. 2b is indicated near the corresponding Hall plateaus ($R_{xy}^{-1}=\pm\nu e^2/h$). We have also measured $R_{xx}$ and $R_{xy}$ vs. *B* at fixed $V_{gate}$ showing also half-integer QHE for both electrons (Fig. 2c) and holes (Fig. 2d). First observed in exfoliated graphene[4,5], half-integer QHE has been observed only very recently in synthesized graphene, including epitaxial graphene grown on SiC[24-27], CVD graphene grown on Ni[12], and now CVD graphene grown on Cu. Observation of such QHE is an important indication that the graphene fabricated by these synthetic (and more scalable) approaches possesses intrinsic graphene properties with electronic quality comparable with graphene exfoliated from graphite.

Fig. 3a shows $\Delta R_{xx}(B)= R_{xx}(B)-R_{xx}(B=0T)$ measured at various *T*'s in device "B". The low *T* (e.g., 1.5 K) magnetoresistance displays two pronounced features that weaken (and eventually disappears) at elevated *T*: 1) reproducible fluctuations, interpreted as the universal conductance fluctuation (UCF)[28]; 2) an overall negative magnetoresistance at low *B*, interpreted as due to weak localization (WL)[28]. Both UCF and WL are mesoscopic quantum transport phenomena resulted from the phase coherence of charge carriers. In particular, WL has been used as a powerful tool to probe carrier transport (esp. scattering processes) and disorder in a variety of graphene samples fabricated by different methods[6, 29-32]. We have fitted (Fig. 3b inset) our experimental data by WL theory developed for graphene[33]. Fig. 3b shows $L_\varphi$ (dephasing length due to inelastic scattering), $L_i$ (elastic intervalley scattering length) and $L_*$ (elastic intravalley scattering lengths) extracted from such fits and plotted vs. *T*. While $L_\varphi$ increases with decreasing *T* and reaches ~0.3 µm at 1.5 K, $L_i$ and $L_*$ are relatively *T*-insensitive. The fact that all these scattering lengths ($L_\varphi$, $L_i$ and $L_*$) are much smaller than the sample size (~3 µm) suggests the dominant scattering source is not the edge, but rather disorder within the sample, such as impurities trapped near graphene or defects in the graphene lattice[2,30,31,33]. It has been pointed out that inter-valley scattering, which requires atomically-sharp disorder (e.g., point defects)[30-33], is essential for WL in graphene. The relatively short $L_i$ (<~150 nm) observed indicates that an appreciable amount of such disorder is present in our sample. The even shorter $L_*$ (<$L_i$) further suggests the presence of additional source of disorder, such as lattice defects larger than atomic scale (e.g. line defects, dislocations, ripples etc)[2,30].

In summary, we have synthesized wafer-scale graphene with dominant monolayer coverage by ambient pressure CVD on Cu. Our transferrable CVD graphene show intrinsic graphene behavior such as half-integer quantum Hall effect[34], and other excellent electronic properties characterized by the ambipolar field effect, carrier mobility and phase coherence. The large, flexible and transferrable graphene films synthesized with a simple and scalable method and possessing excellent uniformity and electronic quality can enable a wide range of applications exploiting the exceptional properties of graphene.

Contributions by H.C. and Q.Y. are equally important to this work. We thank Miller Family Endowment, NRI-MIND, State of Indiana, ACS and UH-CAM for financial support. Part of the work was carried out at the National High Magnetic Field Laboratory, which is supported by NSF and the State of Florida. We thank G. Jones, T.Murphy, J-H.Park and E. Palm for experimental assistance and R.Colby, D.Pandey and E.A. Stach for helpful discussions.

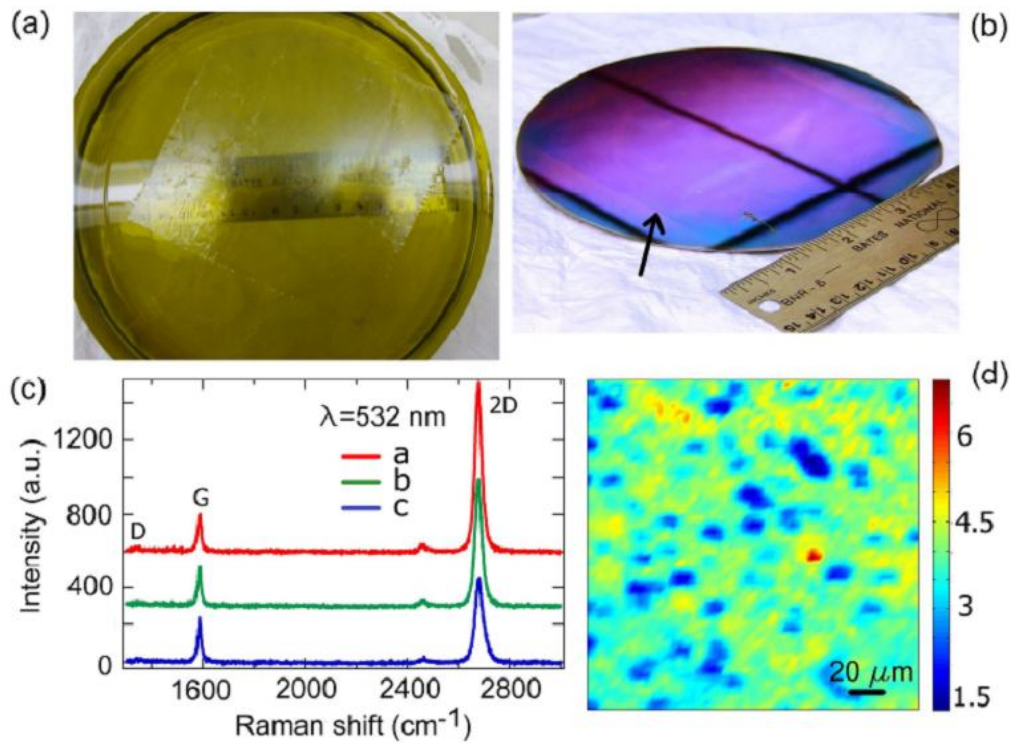

**Fig.1**. Photograph of a 4in× 4in CVD graphene film (a) coated with PMMA and floating on liquid after etching off the Cu substrate and (b) transferred on a Si wafer, with PMMA removed (arrow marks the edge of graphene and thick black lines on the wafer are room ceiling reflections). (c) Representative Raman spectra (c, offset for clarity) measured (with a 532nm laser) in a CVD graphene film transferred to $SiO_2$/Si. The 2D band can all be fitted by a single Lorentzian, with center ~2680 cm$^{-1}$ and FWHM ~34 cm$^{-1}$, consistent with previously observed values (Ref. 15). (d) Raman map of $I_{2D}/I_G$ over a 200 μm ×200 μm area, most (99%) of which can be associated with monolayer ($I_{2D}/I_G$ >2).

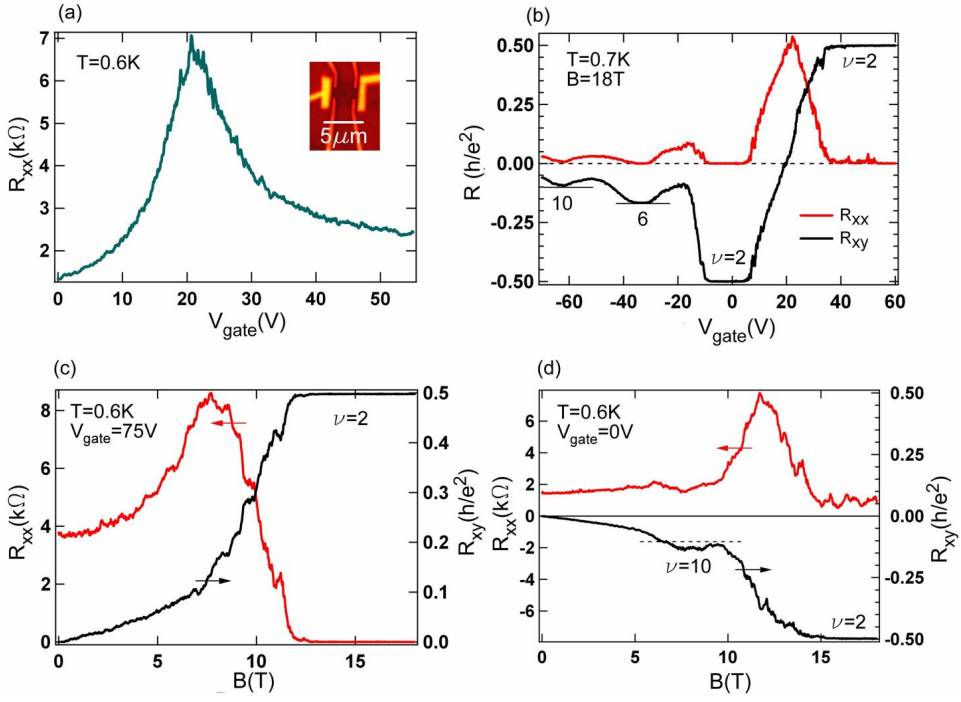

**Fig. 2**. Ambipolar field effect and half integer quantum Hall effect (QHE) of CVD graphene. (a) Four-terminal longitudinal resistance ($R_{xx}$) as a function of gate voltage ($V_{gate}$) measured in device "A" (optical image shown in the inset). (b) $R_{xx}$ and $R_{xy}$ (Hall resistance) as a function of gate voltage at perpendicular magnetic field B=18T and low temperature (T=0.7K). (c), (d) $R_{xx}$ and $R_{xy}$ as functions of B at T=0.6K for $V_{gate}$ =75V (n-type carriers) and $V_{gate}$ =0V (p-type carriers), respectively. The Landau filling factors (ν) of the observed quantum Hall states are labeled in (b-d) and selected QHE plateaus corresponding to $R_{xy} = h/\nu e^2$ are indicated by horizontal lines.

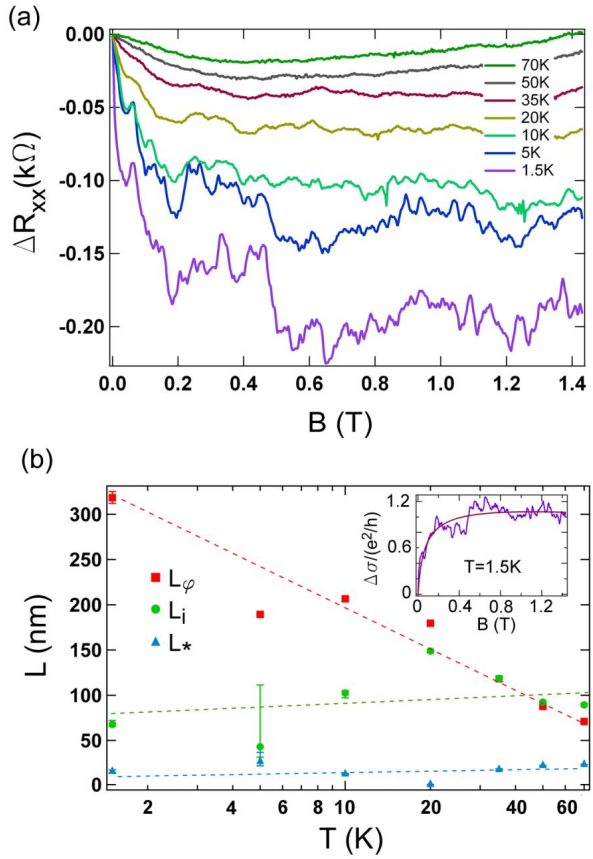

**Fig. 3**. (a) Magnetoresistance $\Delta R_{xx}(B)= R_{xx}(B)-R_{xx}(B=0T)$ measured in device "B" at various temperatures. (b) Extracted characteristic lengths from weak localization as a function of the temperature. Dashed lines are guides to the eye. Inset shows the magnetoconductivity (calculated from measured data and normalized by $e^2/h$) $\Delta\sigma_{xx}(B)= \sigma_{xx}(B)-\sigma_{xx}(0T)$ at T=1.5 K. Solid line is the fit using the WL theory for graphene (Ref. 33).

12